\documentclass[fleqn,twoside]{article}
\usepackage{espcrc2}

\usepackage{graphicx}

\def\bc{\begin{center}}
\def\ec{\end{center}}
\def\beq{\begin{equation}}
\def\eeq{\end{equation}}
\def\bs{\begin{slide}}
\def\es{\end{slide}}
\newcommand{\bmath}{\begin{displaymath}}
\newcommand{\emath}{\end{displaymath}}
\newcommand{\beqn}{\begin{eqnarray}}
\newcommand{\eeqn}{\end{eqnarray}}
\newcommand{\beqns}{\begin{eqnarray*}}
\newcommand{\eeqns}{\end{eqnarray*}}
\newcommand{\ba}{\begin{array}{c}} 
\newcommand{\bat}{\begin{array}{cc}} 
\newcommand{\ea}{\end{array}} 
\newcommand{\lef}{(1-\gamma_5)}

\newcommand{\lsim}{\stackrel{<}{_\sim}}
\newcommand{\gsim}{\stackrel{>}{_\sim}}
\newcommand{\gev}{\, \mbox{GeV}}
\newcommand{\mev}{\, \mbox{MeV}}
\newcommand{\Frac}[2]{\frac{\displaystyle #1}{\displaystyle #2}}

\title{Hadronic decays of the tau lepton~: Theoretical outlook
\thanks{IFIC/07$-$06 report. Talk given at the 9$^{th}$ International Workshop on Tau
Lepton Physics, TAU06, 19th-22nd September 2006, Pisa (Italy).}}

\author{J. Portol\'es \address[IFIC]{Instituto de F\'{\i}sica Corpuscular, IFIC,
                                     CSIC-Universitat de Val\`encia, \\
                                     Edifici d'Instituts de Paterna,
                                     Apt. Correus 22085, E-46071 Val\`encia, Spain}}
       
\begin{document}

\begin{abstract}
The structure of the form factors stemmed from the hadronization of QCD
currents in the energy region of the resonances can be explored through the
analyses of exclusive hadronic decays of the tau lepton.
I give a short review on the later theoretical progress achieved in the description
of experimental data.
\end{abstract}

\maketitle

\section{Introduction}
Our comprehension of the dynamics underlying strong interactions in the hadronic 
low-energy region is entangled due to our lack of knowledge on the implementation
of Quantum Chromodynamics in the non-perturbative regime. In spite of the success
of QCD in the description of strong interactions at large energies, in the 
domain of asymptotic freedom, the study of processes involving the lower part of the
hadronic spectrum (characteristically $E \lsim 2 \mbox{GeV}$) is not feasible with
a strong interaction theory written in terms of dynamical quarks and gluons.
Ideally the way out would be to trade partonic QCD for
a dual theory written in terms of the relevant degrees of freedom, i.e. mesons and
baryons. However we do not (yet) know how to proceed to reach this goal. In 
Ref.~\cite{Portoles:2004vr} I already pointed out different approaches that are 
usually followed in order to describe the phenomenology of hadrons (other than
lattice gauge calculations), namely {\rm ad-hoc} parameterisations,  
phenomenological Lagrangian models and effective field theories. I recall their main 
features~:
\vspace*{0.3cm} \\
{\em Parameterisations} \vspace*{0.1cm} \\ \hspace*{0.3cm} The main goal
is to provide an expression for the amplitudes on account of the
supposedly known dynamics that drives the process~: resonance saturation, polology, etc. 
The simplicity of these parameterisations allows their easy implementation
in the analyses of experimental data, however the connection between the parameters
and QCD is not known. Moreover the initial dynamical assumptions are uncontrolled and
may be in conflict with the underlying theory, therefore very little is learned from
Nature from this approach. 
\vspace*{0.3cm} \\
{\em Models of phenomenological Lagrangians} \vspace*{0.1cm} \\ \hspace*{0.3cm}
Written in terms of hadron fields, phenomenological Lagrangians are driven by
assumptions whose link with QCD is, in many cases, not clear. Well known examples
of these models describing the strong interaction in the presence of resonances are the
Hidden Symmetry or Gauge Symmetry Lagrangians \cite{models} where vector mesons are introduced
as gauge bosons of suggested local symmetries. 
\vspace*{0.3cm} \\
{\em Effective field theories}
\vspace*{0.1cm} \\ \hspace*{0.3cm} Chiral symmetry of massless QCD and its spontaneous breaking
can be used to construct a strong interaction field theory involving the lightest $SU(3)$ octet
of pseudoscalar mesons \cite{chiral,chiral1}. Known as Chiral Perturbation
Theory ($\chi PT$), it has been very much useful in the study of strong interaction effects at very
low energy, namely $E \ll M_{\rho}$ (being $M_{\rho}$ the mass of the $\rho (770)$,
the lightest hadron not included in the theory), and it is an illustration of an effective
field theory (EFT). The latter tries to embody the main features of the fundamental theory in order
to handle this one in a specific energy regime where it is, whether more inconvenient or just
impossible, to apply it \cite{eft}. In order to proceed to the construction of an EFT it is 
necessary the existence of a gap in the spectrum of masses that sets apart those degrees of freedom
to be integrated out from those whose dynamics remain in the theory. In $\chi PT$, for instance, it
is the one separating the lightest octet of pseudoscalar mesons from the light-flavoured resonance
states.
\vspace*{0.3cm} \\
\hspace*{0.3cm}
Semileptonic processes stemmed from the hadronization of QCD currents into exclusive channels
constitute singular physical systems where the study of non-perturbative strong dynamics is
easier than in pure non-leptonic processes. This is so because, on one
side, lepton and hadron sectors factorise cleanly and, moreover, exclusive channels give valuable
information on the dynamics of the interaction itself, hence on the realization of the
non-perturbative strong interaction in this energy region. In addition a good deal of information
is known about form factors of QCD currents from different model-independent sources such as 
parton dynamics, analyticity or unitarity. The latter also pervade general amplitudes but their
application on the form factors is much simpler.
\par
The tau lepton, $M_{\tau}= \left(1776.99^{+0.29}_{-0.26}\right) \mbox{MeV}$ 
\cite{Yao:2006px}, decays hadronically in an energy region populated by light-flavoured resonances.
Hence exclusive semileptonic tau decays offer a unique setting where to explore resonance
dynamics through the form factors arisen in the hadronization of QCD currents.
Within the Standard Model the matrix amplitude for the exclusive hadronic decays of the tau 
lepton, $\tau^{-} \rightarrow H^{-} \nu_{\tau}$, is generically given by
\beq
{\cal M} \, = \, \Frac{G_F}{\sqrt{2}} \,  
V_{\tiny{CKM}} \, \overline{u}_{\nu_{\tau}} \gamma^{\mu} 
\lef u_{\tau} \, {\cal H}_{\mu} \, , 
\label{eq:m}
\eeq
where
\beq
{\cal H}_{\mu}\,  = \, \left\langle \, H \,\left| \,\left( {\cal V}_{\mu} - {\cal A}_{\mu}\right) \,
e^{i {\cal L}_{QCD}} \,\right| \, 0 \, \right\rangle \; ,
\label{eq:h}
\eeq
is the hadron matrix element of the left current (notice that it has to be evaluated 
in the presence of the strong interactions driven by ${\cal L}_{QCD}$). Symmetries help
to define a decomposition of ${\cal H}_{\mu}$ in terms of the allowed Lorentz structures
of implied momenta and a set of functions of Lorentz invariants, the {\em hadron form
factors} $F_i$ of QCD currents, 
\beq
{\cal H}_{\mu} \, = \,  \,  \sum_i \! \! \underbrace{ \;   ( \,  \, \ldots \, \, )_{\mu}^i 
\; }_{Lorentz \, structure} \! \! \! \! \! \! 
 F_i (q^2, \ldots) \; .
\label{eq:ff}
\eeq
\par
An analogous discussion of the hadronic decays of the tau lepton can be carried out
in terms of the {\em structure functions} $W_X$ defined in the
hadron rest frame \cite{KS2}~:
\beq
d\Gamma  =  \frac{G_F^2}{4 M_\tau} \, |V_{CKM}|^2 \, {\cal L}_{\mu\nu} \, {\cal H}^{\mu} \,
{\cal H}^{\nu*} \, dPS \;  ,
\eeq
with
\beq
{\cal L}_{\mu\nu} \, {\cal H}^{\mu} \, 
{\cal H}^{\nu*}  =  \sum_X \, L_X \, W_X \; ,
\label{eq:sff}
\eeq
where ${\cal H}_\mu$ is the hadronic current in Eq.~(\ref{eq:h}), ${\cal L}_{\mu\nu}$
carries the information of the lepton sector and $dPS$ collects the appropriate phase
space terms. Structure functions can be written in terms of the relevant form factors
and kinematical components. They contain the dynamics of the hadronic decay and
their reconstruction can be accomplished through the study of spectral functions or 
angular distributions of data.
The number of structure functions depends,
clearly, of the number of hadrons in the final state. For a two-pseudoscalar case there
are 4 of them. For a three-pseudoscalar process the total number of structure functions
is 16.
\par
I will focus in the decays of the tau lepton into a few channels with two and three
pseudoscalars. 
The theoretical description of the dynamics that drives the decays into
more than three pseudoscalars still relies in the model-independent isospin counting \cite{pppp}.

\section{Breit-Wigner parameterisations}
As already commented, strong interaction dynamics in hadronic tau decays involves the
resonance energy region, hence it is driven by those states.  
This is the well known concept of {\em resonance dominance}
that has pervaded hadron dynamics since the first stages of the study of the strong 
interaction. It is a widespread common lore that resonances should be
functionally described by Breit-Wigner parameterisations. While it is clear that polology
demands this structure, its connection with QCD is still lacking. In fact it has already
been proven, for instance, that the description of hadronic tau decays through them is
not consistent with the chiral symmetry of massless QCD \cite{joVictoria,us3pion}.
\par
Experiments like ALEPH, CLEO, DELPHI and OPAL 
\cite{exp0,exp1,exp15,exp2,exp21,exp24,exp25,exp3} have
collected an important set of experimental data on
hadronic decays of the tau lepton into exclusive channels. In the last two years both
BABAR and BELLE experiments have joined in this effort \cite{babar,belle1,belle2}, and the
prospects for a SuperB factory are very much promising \cite{superB}.
Analyses of these data are usually carried out using the TAUOLA library \cite{tauola} that includes
parameterisations of the hadronic matrix elements. At present the latter only includes 
Breit-Wigner specifications for the form factors.
Its application to the hadronization of charged QCD currents in tau decays has a 
long story \cite{preKS,KS1} that boils down into a series of articles
\cite{KS3,KS5,KS4} that carry an exhaustive analysis of the tau decays up to three pseudoscalars.
\par
The presently employed parameterisation, the Generalized K\"uhn-Santamar\'{\i}a model (GKS),
is obtained by combining Breit-Wigner factors ($BW_R(q^2)$), in general non-linearly, according to 
the expected resonance dominance in each channel, for instance,
\begin{equation}
F(q^2) = {\cal N} \, \sum_i \, \alpha_i \, BW_{R_{i}}(q^2) \; ,
\end{equation} 
where $ {\cal N} $ is a normalisation required to fulfill the chiral symmetry expansion at
${\cal O}(p^2)$.  
Then data are analysed by fitting the $\alpha_i$ parameters and those present in the Breit-Wigner
factors (masses, on-shell widths). Two different functions are applied~:
\vspace*{0.2cm} \\
a) {\em K\"uhn-Santamar\'{\i}a Model} \\
The Breit-Wigner factors are given by \cite{preKS,KS1}
\begin{equation}
BW_{R_i}^{KS}(s) = \Frac{M_{R_i}^2}{M_{R_i}^2 -s-i \, \sqrt{s} \, \Gamma_{R_{i}}(s)}  \, ,
\label{eq:KS}
\end{equation} 
that guarantees the right asymptotic behaviour, ruled by QCD, for the form factors. 
\vspace*{0.2cm} \\
b) {\em Gounaris-Sakurai Model } \\
Originally constructed to study the role of the $\rho(770)$ resonance in the vector form
factor of the pion \cite{GS}, its use has been
extended to other hadronic resonances \cite{exp0,exp1,KS4}. The Breit-Wigner function
now reads~:
\begin{equation}
BW_{R_i}^{GS}(s) = \Frac{M_{R_i}^2+ f_{R_i}(0)}{M_{R_i}^2-s+f_{R_i}(s)-i \sqrt{s} \, \Gamma_{R_i}(s)} \, ,
\label{eq:GS}
\end{equation} 
where $f_{R_i}(s)$ carries information on the specific dynamics of the resonance 
and $f_{\rho(770)}(s)$ can be read from Ref.~\cite{GS}. 
\vspace*{0.2cm}\\
The procedure applied by the experimental 
groups when using these parameterisations \cite{exp0,exp1} is to regard both models and consider the
discrepancy between them as an estimate of the theoretical error.
\vspace*{-0.1cm}
\section{Effective Theory like model~:  Resonance Chiral Theory}
\label{sect:phenol}
At variance with $\chi PT$,
the lack of a mass gap between the spectrum of light-flavoured meson resonances and the perturbative
continuum (let us say $E \gsim 2 \, \mbox{GeV}$) prevents the construction of an appropriate EFT to handle
the strong interaction in the energy region spanned by tau decays. However there are several tools that
allow us to grasp relevant features of QCD and to implement them in an EFT-like Lagrangian
model. The two key premises are the following~:    
\par
1) A theorem put forward by S.~Weinberg \cite{chiral} and worked out by H.~Leutwyler \cite{leut} 
states that, if one writes down the most general possible Lagrangian, including all terms
consistent with assumed symmetry principles, and then calculates matrix elements with this
Lagrangian to any given order of perturbation theory, the result will be the most general
possible S-matrix amplitude consistent with analyticity, perturbative unitarity, cluster decomposition
and the principles of symmetry that have been specified.
\par
2) It has been suggested \cite{ncc} that the inverse of the number of colours of the
gauge group $SU(N_C)$ could be taken as a perturbative expansion parameter. Indeed 
large-$N_C$ QCD shows features that resemble, both qualitatively and quantitatively, the
$N_C=3$ case. Relevant consequences of this approach are that meson dynamics in the 
large-$N_C$ limit is described by the tree diagrams of an effective local Lagrangian;
moreover, at the leading order, one has to include the contributions of the infinite
number of zero-width resonances that constitute the spectrum of the theory.
\par
Both assertions can be merged by constructing a Lagrangian theory in terms of $SU(3)$ 
(pseudoGoldstone mesons) and $U(3)$ (heavier resonances) flavour multiplets as active degrees of freedom.
This has systematically
been established \cite{vmd1,rcht2,Cirigliano:2006hb} and sets forth the following features~: \vspace*{0.2cm} \\
i) The construction of the operators is guided by chiral symmetry for the lightest pseudoscalar
mesons and by unitary symmetry for the resonances. Typically,
\begin{equation} \label{eq:operator}
{\cal O} \, \sim \; \langle \, R_1 R_2 \dots \chi (p^n) \, \rangle \; ,
\end{equation}
where $R_i$ indicates a resonance field and $\chi (p^n)$ is a chiral structured tensor,
involving the Goldstone bosons, with a chiral counting represented by the power of the momenta.
Then chiral symmetry is preserved upon 
integration of the resonance states and the low-energy expansion of the amplitudes show the 
appropriate behaviour. \vspace*{0.2cm} 
\\
ii) As in $\chi PT$, symmetries do not provide information on the weights of the operators, i.e. on their
coupling constants. The latter only incorporate information from higher energies and, in principle, are
completely unknown. However if we want to disguise our theory with the role of mediator between the chiral and 
the parton regimes it is clear that the amplitudes or form factors arising from our Lagrangian have to
match the asymptotic behaviour driven by QCD. A heuristic strategy, well supported at the 
phenomenological level \cite{greenf,Cirigliano:2006hb}, resides in matching the OPE of Green functions
(that are order parameters of chiral symmetry) with the corresponding expressions evaluated within
the theory. This procedure provides a determination for some of the coupling constants of the Lagrangian.
In addition the asymptotic behaviour of form factors of QCD currents is estimated from the spectral
structure of two-point functions \cite{Rosell:2007kc} or the partonic make-up \cite{Brodsky}.
\vspace*{0.2cm} \\
iii) The theory that we have devised, known as Resonance Chiral Theory ($R \chi T$), lacks an expansion
parameter. There is of course the guide given by $1/N_C$ that translates into a loop expansion. 
However there is no counting that limits the number of operators with resonances that have to be included in the
initial Lagrangian. This non-perturbative character of $R \chi T$, that may take back the perturbative practitioner,
has to be understood properly. On one side the number of resonance fields relies fundamentally in the
physical system of interest, on the other, the maximum order of the chiral tensor in Eq.~(\ref{eq:operator}) is 
very much constrained by the enforced high-energy behaviour, as explained in ii) before. Customarily 
higher powers of momenta lean to spoil the asymptotic conduct that QCD demands \cite{rcht2,Cirigliano:2006hb}.
Therefore there is a well defined methodology to deal with $R \chi T$. 
\vspace*{0.2cm} \\
As commented above large-$N_C$ requires, already at $N_C \rightarrow \infty$,  an infinite spectrum in order to 
match the leading QCD logarithms. At present $R \chi T$ only includes one multiplet of resonances for the 
different quantum numbers~: scalars, pseudoscalars, vectors and axial-vectors. It is not known how to include
an infinite number of states in a model-independent way and this simplification can produce inconsistencies in the
matching procedure described above \cite{Bijnens:2003rc}. In principle a way out of these lean on the inclusion 
of more states that may delay the appearance of that problem. From a phenomenological point of view, though, 
the first multiplets drive the relevant dynamics in the systems of interest, as hadronic tau decays,
and are clearly enough. However there is no conceptual problem that prevents the 
addition of more spectra (in a finite number) if needed.  

\section{Hadronic off-shell widths of meson resonances}
The hadronic decays of the tau lepton happen in an energy region where resonances
do indeed resonate. Therefore the leading large-$N_C$ prescription of zero-width resonances
does not allow to perform a phenomenological study of the decays. The introduction of finite widths, 
in Eqs.~(\ref{eq:KS},\ref{eq:GS}), should be done through the same tools used to handle the amplitudes. 
For narrow resonances, like most of those with $I=0$ in the energy region
spanned by tau decays, it is a good approximation to consider constant widths that can be
taken from the phenomenology at hand or fitted from data. Wider resonances, though, have a non-trivial
off-shell structure that has to be taken into account.  
\par
The off-shell width of the $\rho(770)$ has been studied thoroughly and it is dominated by 
the $\pi \pi$ contribution. In the GKS
parameterisations the imaginary part of the mass in the pole reads \cite{GS}~:
\beq
\sqrt{s}  \, \Gamma_{\rho}(s)  =  \Gamma_{\rho}(M_{\rho}^2)   \Frac{s}{M_{\rho}} 
\frac{\sigma^3(s)}{\sigma^3(M_\rho^2)}  \, \theta \left( s-4M_{\pi}^2 \right)  ,
\label{eq:wGS}
\eeq
where $\sigma(s) = \sqrt{ 1-4M_{\pi}^{2}/s }$. 
In Ref.~\cite{widthro} it was seen that this width can be evaluated within $R \chi T$
through a Dyson-Schwinger like resummation controlled by the short-distance behaviour
required by QCD on the correlator of two vector currents. The result for the imaginary 
part of the pole is~:
\beq
M_{\rho} \, \Gamma_{\rho}(s) \, = \, \Frac{M_V^2 \, s}{96 \, \pi \, F^2} \,
 \sigma^3(s) \, \theta \left( s-4M_{\pi}^{2} \right)  \, ,
\label{eq:wUS}
\eeq
were $M_V$ is the vector resonance mass and $F$ the decay constant of the pion, both
of them in the chiral limit. 
It is worth to notice that the dependence on the $s$ variable of both imaginary parts
in Eqs.~(\ref{eq:wGS},\ref{eq:wUS}) is the same. However the prescription given in 
Eq.~(\ref{eq:wUS}) already accounts for the on-shell width while in the GKS model it is 
a free parameter to be fitted.
\par
The off-shell width of $K^*(892)$ follows from that on $\rho(770)$. However in 
Ref.~\cite{Jamin:2006tk} it was concluded that it requires a slight modification~:
\begin{equation} \label{eq:gkss}
\Gamma_{K^*}(s) = \Frac{G_V^2 M_{K^*} s}{64 \pi F^4}  \sigma_{K \pi}^3 
\frac{D_1(r q_{K \pi}(M_{K^*}^2))}{D_1(r q_{K \pi}(s))} \; 
\end{equation}
for $s >(M_K + M_{\pi})^2$, where $q_{K \pi}(s)$ is the kaon momentum in the rest frame
of the hadronic system, $\sigma_{K \pi} = 2 q_{K \pi}/\sqrt{s}$ and $D_1(x) = 1 + x^2$ is a Blatt-Weisskopf barrier factor with
$r$ the interaction radius. We find $r=(3.5 \pm 0.6) \, \mbox{GeV}^{-1}$ and 
$G_V = (64.6 \pm 0.4) \mev$  \cite{Jamin:2006tk}, when the $K \eta$ contribution to the width in 
Eq.~(\ref{eq:gkss}) is also included.
\par
The hadronic off-shell widths for other resonances like $\rho(1450)$ or $a_1(1260)$, that
are also relevant in the decays of the tau lepton, are not so well known. In principle
the methodology put forward in Ref.~\cite{widthro} could also be applied but it is necessary
to know better the perturbative loop expansion of $R \chi T$ in order to proceed. Therefore
one has to resort to appropriate modelizations being the key point the leading
behaviour of the off-shell structure in the $s$ variable. Hence it is customary to 
propose parameterised widths of which the simplest version reads as~:
\beq
\Gamma_R(s) = \Gamma_R(M_R^2) \Frac{\Phi(s)}{\Phi(M_R^2)} \left( \Frac{M_R^2}{s} \right)^{\alpha} 
\theta(s-s_{th}),
\label{eq:wa1}
\eeq
where the $\Phi(s)$ function is related with the available phase-space that corresponds to 
the threshold given by $s_{th}$. The $\alpha$ parameter can be given by models or fitted to
experimental data. This last procedure was used in Ref.~\cite{us3pion} to obtain information
on the $a_1(1260)$ width from the tau decay into three pions, giving $\alpha \sim 5/2$. A 
thorough study on the off-shell widths of resonances that QCD demands is still missing.

\section{$\tau^- \rightarrow \pi^- \pi^0 \nu_{\tau}$~: Vector form factor}
The vector form factor of the pion, $F_V(s)$, is defined by~:
\beq
\left\langle  \pi^+(p') \, \pi^-(p)  \left| \, {\cal V}_{\mu}^{3} \, \right| 
 0  \right\rangle 
 =   \, \left( p - p' \right)_\mu F_V(s)   ,
\eeq
where $s=(p+p')^2$ and ${\cal V}_{\mu}^3$, the third component of the vector current
associated to the $SU(3)$ flavour symmetry of the QCD Lagrangian.  This form factor drives
the isovector hadronic part of $e^+ e^- \rightarrow \pi^+ \pi^-$ and, in the isospin limit,
of $\tau^- \rightarrow \pi^- \pi^0 \nu_{\tau}$.
At very low energies, $E \ll M_{\rho}$, $F_V(s)$ has been studied in the $\chi PT$ framework up to 
${\cal O}(p^6)$ \cite{op6fv,cfu1,cfu2}. Here we collect the last relevant developments.
\vspace*{0.2cm} \\
a) \underline{$M_{\rho} \lsim E \lsim 1 \gev$}
\vspace*{0.1cm} \\
This energy region is dominated by the $\rho(770)$ and, accordingly, its study is
relevant to determine the parameters of this resonance. In addition it gives the largest
contribution to the hadronic vacuum polarisation piece of the anomalous magnetic moment
of the muon \cite{g-2,ty}.
\par
The authors of Ref.~\cite{gp97}
proposed a framework where the ${\cal O}(p^4)$ $\chi PT$ low-energy result is matched
 at higher energies with an expression driven by vector meson dominance that is modulated by an
Omn\`es solution of the dispersion relation satisfied by the vector form factor
of the pion. It provides an excellent description of the $\rho(770)$ up to energies of $1 \gev$.
The more involved procedure of the unitarization approach \cite{oop} gives also a good description
of this energy region.
\par
A model-independent parameterisation of the vector form factor constructed on grounds of 
an Omn\`es solution for the dispersion relation has also been considered \cite{ty,pp01,pp02}. 
This approach can be combined with $R \chi T$ \cite{pp01} and it is able to give some improvement
over the previous approach if one includes
information on the $\rho(1450)$ through the $\pi \pi$ elastic phase-shift input in the 
Omn\`es solution. Hence it extends the description of the form factor up to $E \sim 1.3 \gev$.
Our analysis gets $M_{\rho} = (775.9 \pm 0.5) \mev$.
\par
\begin{figure}[t]
\begin{center}
\hspace*{-0.3cm}\includegraphics*[scale=0.331]{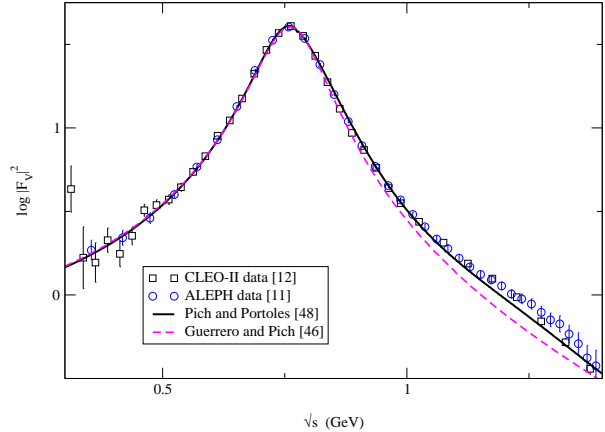}
\vspace*{-1.2cm}
\caption[]{\label{fig:1} Comparison of the vector form factor of the pion as given by
tau data and the theoretical description of Refs. \cite{gp97,pp01}.}
\vspace*{-1cm}
\end{center}
\end{figure}

A comparison of the theoretical descriptions given by Refs.~\cite{gp97,pp01} and the 
experimental data by ALEPH \cite{exp0} and CLEO \cite{exp1} is shown in Figure~\ref{fig:1}. 
A new collection of data, still not analysed within the above mentioned frameworks, has recently
been provided by CMD-2 \cite{Akhmetshin:2006bx}. Also final state interactions in KLOE data 
\cite{Aloisio:2004bu} have been studied with an {\em ad hoc} parameterization \cite{Pancheri:2006cp}.

\vspace*{0.1cm}
\noindent
b) \underline{$1 \gev \lsim E \lsim 2 \gev$}
\vspace*{0.1cm} \\
The extension of the description of the vector form factor of the pion at higher energies
involves the inclusion of further information. Up to $2 \gev$ two $\rho-$like resonances
play the main role~: $\rho(1450)$ and $\rho(1700)$. However the interference between resonances,
the possible presence of a continuum component, etc. still deserve a study not yet done.
\par
The inclusion of $\rho(1450)$ only improves
slightly the behaviour when a Dyson-Schwinger-like resummation is performed in the
framework of $R \chi T$ \cite{rchloop1}. Lately, and based in a previous modelization proposed in 
Ref.~\cite{do01}, a procedure to extend the description of the vector form factor of 
the pion at higher energies has been put forward \cite{KS4}. The proposal for the 
form factor embodies
a Breit-Wigner parameterisation using the GKS model
to describe $\rho(770)$, $\rho(1450)$ and $\rho(1700)$ resonances, appended with a
modelization of large-$N_C$ QCD which sums up an infinite number of zero-width 
resonances to yield a Veneziano type of structure.
\begin{figure}[t]
\begin{center}
\vspace*{-0.2cm}
\includegraphics*[scale=0.48]{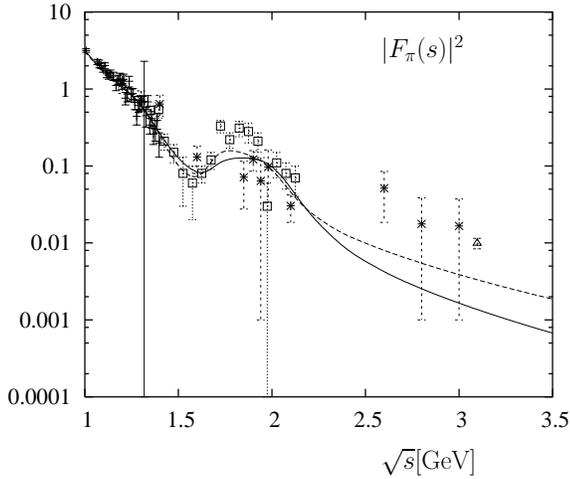}
\vspace*{-1cm}
\caption[]{\label{fig:2} Comparison of the vector form factor of the pion in the energy
region above $1 \gev$ from Ref.~\cite{KS4}. Solid (dashed) line corresponds to the
parameterisation given in Eq.~(\ref{eq:KS}) (Eq.~(\ref{eq:GS})).}
\vspace*{-0.9cm}
\end{center}
\end{figure}
 In Figure~\ref{fig:2} it is shown how
this parameterisation compares with data. The description is reasonable up to $2 \gev$.
Above this region there is almost no data though, in principle, it looks quite compatible
with it. This model has been recently employed to analyse new Belle data \cite{Abe:2005ur}
and it is claimed that data shows sensitivity to the $\rho(1700)$ resonance.
\par
The key role that plays the vector form factor of the pion in the hadronic 
vacuum polarisation contribution to the anomalous magnetic moment of the pion \cite{g-2},
together with the seeming discrepancy between the predictions provided by 
$e^+ e^- \rightarrow \pi^+ \pi^-$ \cite{Akhmetshin:2006bx,Aloisio:2004bu}
 and $\tau^- \rightarrow \pi^- \pi^0 \nu_{\tau}$
data, set up the issue of the size of isospin violation. A thorough analysis of the radiative
corrections in $\tau^- \rightarrow \pi^- \pi^0 \nu_{\tau}$ and other relevant isospin
violating sources (kinematics, short-distance electroweak corrections, $\rho-\omega$ mixing)
was carried out in Ref.~\cite{Vincenzo}. Recently it has also been stated that further 
model-dependent resonance contributions not taken into account before are also relevant
\cite{Flores-Baez:2006gf}, in particular the seemingly large coupling $\rho \omega \pi$,
and may correct significantly the decay rate. Additional evaluations along this claim should
be done in order to reach a more sounded conclusion.

\section{$\tau \rightarrow K \pi \nu_{\tau}$~:  Vector and scalar form factors}

The dominant contribution to the Cabibbo-suppressed tau decay rate is due to
the decay $\tau \rightarrow K \pi \nu_{\tau}$. The corresponding distribution
function has been measured experimentally in the past by ALEPH \cite{Barate:1999hj} 
and OPAL \cite{Abbiendi:2004xa}. With the large data sets on hadronic tau decays from the
B-factories both BABAR and Belle are, at present, analysing their data \cite{belle1}.
\par
\begin{figure}[t]
\begin{center}
\vspace*{-0.2cm}
\includegraphics*[scale=0.3]{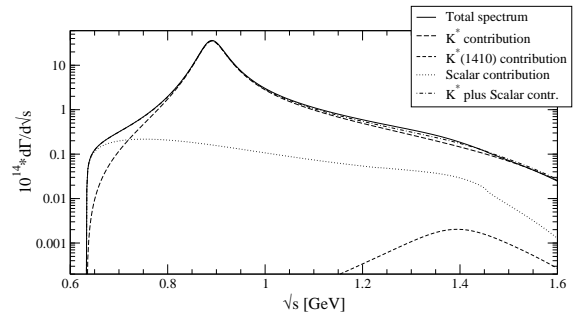}
\vspace*{-0.9cm}
\caption[]{\label{fig:3} Differential decay distribution of the decay 
$\tau \rightarrow K \pi \nu_{\tau}$
and the different individual contributions computed in Ref.~\cite{Jamin:2006tk}.}
\vspace*{-0.9cm}
\end{center}
\end{figure}

Assuming isospin invariance the relevant hadronic matrix element is guided by two 
form factors~:
\begin{eqnarray}
\langle \,  \pi(p) \,  K(p') \, | \, \overline{s} \, \gamma_{\mu} \, 
 u \, | \, 0 \, \rangle \,  = \,  \nonumber \\
\left( g_{\mu \nu} - \frac{Q_{\mu} Q_{\nu}}{Q^2} \right) \, \left( p' - p \right)^{\nu}  \, 
F_{+}^{K \pi}(Q^2) \, \nonumber \\
+ \, (p' + p)_{\mu} \, \frac{\Delta_{K\pi}}{Q^2} \, F_0^{K \pi}(Q^2) \, ,
\end{eqnarray}
where $Q_{\mu} = (p+p')_{\mu}$, $\Delta_{K \pi} = M_K^2 - M_{\pi}^2$, and 
$F_+^{K \pi}$ ($F_0^{K \pi}$) is the vector (scalar) form factor. These have been studied 
within the GKS model \cite{Finkemeier:1996dh}. A more thorough analysis has been carried
out recently \cite{Jamin:2006tk} where both form factors have been constructed according to
the following settings~: \\
1) {\em Scalar form factor}. We have introduced the meticulous construction given in 
Ref.~\cite{Jamin:2001zq}, where $F_0^{K \pi}$ was determined from $K \pi$ scattering data
by solving the corresponding dispersion relations in a coupled channel formalism. \\
2) {\em Vector form factor}. Following the methodology put forward in Ref.~\cite{gp97} we 
have constructed $F_+^{K \pi}$ by demanding it satisfies both correct chiral limit and
appropriate asymptotic behaviour. We have included explicitly $K^*(892)$ and $K^*(1410)$
with a free parameter ($\gamma$) that weights the contribution of the later resonance,
obtaining from the total branching ratio $\gamma = 0.013 \pm 0.017$ . \\
In Figure~\ref{fig:3} we show the differential decay distribution of the decay and the
different contributions. Notice the fundamental role played by the scalar form factor
at threshold though its contribution to the branching is tiny ($\sim 10^{-4}$).
Moreover we obtain $B[\tau \rightarrow K^*(892) \nu_{\tau}] = 
(1.253 \pm 0.078) \%$ to be compared with the PDG value $(1.29 \pm 0.05) \%$ 
\cite{Yao:2006px}. 

Between the different structure functions defined in Eq.~(\ref{eq:sff}) the one 
known as $W_{SG}$ measures the imaginary part of the interference of both form
factors and requires non-trivial phases of the amplitudes that are essential
for an observation  of possible CP violation effects in the difference of 
$W_{SG}$ for the $\tau^+$ and $\tau^-$ decays~:
\begin{equation}
W_{SG} = -2 \frac{\Delta_{K \pi}}{\sqrt{Q^2}} \, q_{K  \pi} \, 
\mbox{Im} \, \left[ F_{+}^{K \pi} \, \left(F_{0}^{K \pi} \right)^* \right] \, . 
\end{equation}

\begin{figure}[t]
\begin{center}
\vspace*{-0.2cm}
\includegraphics*[scale=0.3]{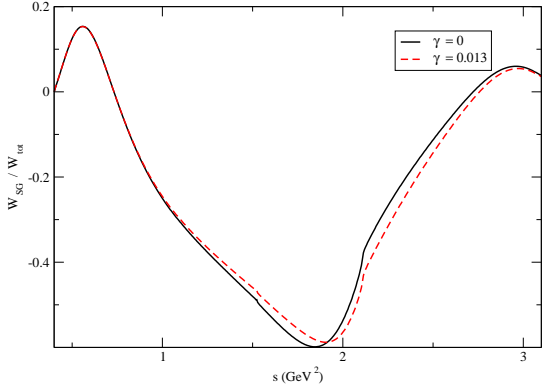}
\caption[]{\label{fig:4} $W_{SG}$ structure function (normalized to $W_{tot}$) for
$\tau \rightarrow K \pi \nu_{\tau}$. $\gamma = 0$ corresponds to exclude the 
$K^*(1410)$ resonance. }
\vspace*{-0.9cm}
\end{center}
\end{figure}

In Figure~\ref{fig:4} we show our result for the form factors given in 
Ref.~\cite{Jamin:2006tk} normalized to~:
\begin{equation}
W_{tot} = W_B \, + \, \frac{3 M_{\tau}^2}{2 Q^2+M_{\tau}^2} \, W_{SA} \, \propto \, 
\frac{d \Gamma}{d Q^2} \; .
\end{equation}
The experimental determination of this structure function requires, however, to
measure full tau kinematics (its direction in the laboratory frame) or to handle a
polarized tau.

\section{$\tau^- \rightarrow (PPP)^- \nu_{\tau}$~:   Vector and axial-vector form factors}
The hadronic matrix element that governs the decay of the tau lepton into three pseudoscalars is parameterised
 by four form factors $F_i$ defined as~:
\begin{eqnarray} \label{eq:fff}
\left\langle P(p_1) P(p_2) P(p_3) \left| \, \left( {\cal V}^{\mu} 
- {\cal A}^{\mu} \right) \, \right| 0 \right\rangle \, = \nonumber \\ \\  
V^{\mu}_1  F_1^A(Q^2,s_1,s_2)\,  + \, V^{\mu}_2 F_2^A(Q^2,s_1,s_2) \nonumber \\ \nonumber \\ 
+ \, Q^{\mu} \, F_3^A(Q^2,s_1,s_2) \, + \,  i \, V^{\mu}_3 F_4^V(Q^2,s_1,s_2) \, , \nonumber
\end{eqnarray} 
where
\begin{eqnarray}
V^{\mu}_1 &=& \left( g^{\mu\nu} - \frac{Q^{\mu}Q^{\nu}}{Q^2}\right)  (p_1-p_3)_{\nu} \, ,\nonumber \\
V^{\mu}_2 &=& \left( g^{\mu\nu} - \frac{Q^{\mu}Q^{\nu}}{Q^2}\right)  (p_2-p_3)_{\nu} \, ,\nonumber \\
V^{\mu}_3 & = & \varepsilon^{\mu\alpha\beta\gamma} p_{1 \alpha} p_{2\beta} p_{3\gamma} \, ,\nonumber \\
Q^{\mu} & = & p_1^{\mu} + p_2^{\mu} + p_3^{\mu} \, , \\
s_i & = & \left( Q-p_i \right)^2 \nonumber \; .
\end{eqnarray}
Here $F_i, i=1,2,3$ correspond to the axial-vector current while $F_4$ drives the vector current.
In the particular case of three pions, we have, due to Bose-Einstein symmetry, 
that $F_2^A(Q^2,s_1,s_2) = F_1^A(Q^2,s_2,s_1)$.
The scalar form factor $F_3^A$ vanishes with the mass of the Goldstone
boson (chiral limit) and, accordingly, gives a tiny contribution in the three pion case.
Finally the vector current only contributes, for the three-pion case, if isospin symmetry
is broken as demanded by G parity conservation; hence in the isospin limit $F_4^V=0$ for this channel.
It gives, in general, a non-vanishing contribution, for other final states. 

\begin{figure*}[!th]
\hspace*{0.5cm}
\includegraphics*[angle=-90,scale=0.6,clip]{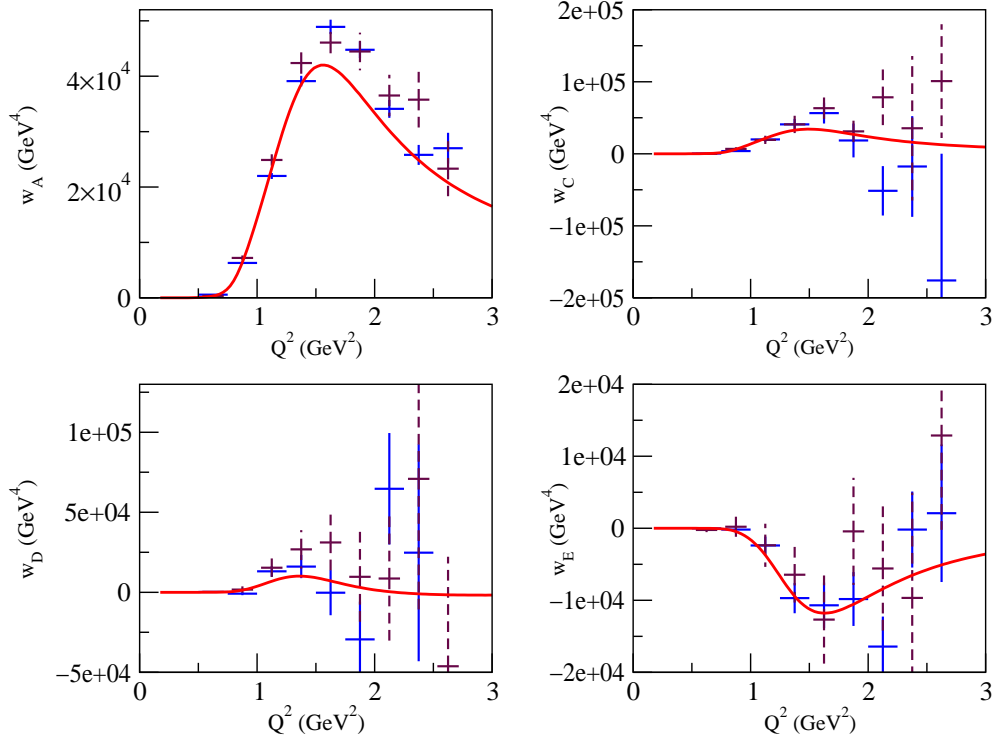}
\vspace*{-0.9cm}
\caption{\label{fig:5} Theoretical predictions for the $w_A$, $w_C$, $w_D$
and $w_E$ integrated structure functions \cite{us3pion} for $\tau \rightarrow \pi \pi \pi \nu_{\tau}$
in comparison with the experimental
data from CLEO-II (solid) \cite{exp24} and OPAL (dashed) \cite{exp2}.}
\vspace*{-0.5cm}
\end{figure*}

\subsection{$\tau^- \rightarrow (\pi \pi \pi)^- \nu_{\tau}$}
As just commented the dynamics of $\tau \rightarrow \pi\pi\pi\nu_{\tau}$ is only driven by 
axial-vector form factors (in the isospin limit) and, accordingly, by the presence of the
axial-vector $a_1(1260)$ modulated by the $I=1$ vector resonances $\rho(770) (\rho)$,
$\rho(1450) (\rho')$ and $\rho(1700) (\rho'')$.
In the very low-energy regime the chiral constraints where explored in Ref.~\cite{FW80}
and later it was calculated up to ${\cal O}(p^4)$  in $\chi PT$ \cite{cfu1}. At higher
energies resonances participate explicitly. In the GKS model the spin 1 axial-vector
form factor is given by~:
\begin{eqnarray}
F_1^A(Q^2, s_i) = {\cal N}|_{\chi {\cal O}(p^2)} \, BW_{a_1}(Q^2) \, \times \nonumber \\ \\
\times \; 
\Frac{BW_{\rho}(s_i) + \alpha \, BW_{\rho'}(s_i) \, + \, \beta \, 
 BW_{\rho''} (s_i)}{1+\alpha+\beta} \; .
\nonumber
\end{eqnarray}
This description, complemented with an {\em ad hoc} construction of the off-shell width 
of the $a_1(1260)$ resonance, provides a good description of the spectrum of three pions
\cite{KS1} though a slight discrepancy shows up in the integrated structure functions \cite{exp24}.
The fit to the data gives the values of the $\alpha$ and $\beta$ parameters,
that compute the weight of each $\rho$-like resonance and, in addition, one can study
masses and on-shell widths of the participating resonances. The issue of isospin violation
in this channel, within the GKS model, has also been considered \cite{isocp}. 
Lately it was shown that this Breit-Wigner parameterisation is not consistent with 
chiral symmetry at ${\cal O}(p^4)$ and thus with QCD \cite{joVictoria,us3pion}.
\par
A thorough study of the 
axial-vector form factors in $\tau \rightarrow \pi\pi\pi\nu_{\tau}$ has been performed in 
Ref.~\cite{us3pion} using the methodology described in Section~\ref{sect:phenol}. 
The authors get a parameterisation of the three pion decay of the tau lepton in terms
of four free parameters~: $M_{a_1}$, $\Gamma_{a_1}(M_{a_1})$, one combination of coupling
constants of the Lagrangian and, finally, the $\alpha$ parameter in the off-shell width of the
$a_1(1260)$ resonance in Eq.~(\ref{eq:wa1}). Next
an analysis of the ALEPH data \cite{exp15} on the spectrum and branching ratio
of $\tau^- \rightarrow \pi^+ \pi^- \pi^- \nu_{\tau}$ is performed, obtaining $M_{a_1} = (1.204 \pm 0.007) \gev$
and $\Gamma_{a_1}(M_{a_1}^2) = (0.48 \pm 0.02) \gev$ where the errors, given by the 
minimisation program, are only statistical.
\par
OPAL \cite{exp2} and CLEO \cite{exp24} have
collected data on the dominant structure functions in the 
$\tau^- \rightarrow \pi^- \pi^0 \pi^0 \nu_{\tau}$ decay, namely, $W_A$, $W_C$,
$W_D$ and $W_E$ (\ref{eq:sff}) that drive the contribution of the $J=1^+$ amplitude
into the process and therefore the {\em integrated structure 
functions} over all the available phase space, defined as~:
\beq
w_{A,C} \, = \, \int \, ds_1 \, ds_2 \, W_{A,C}  \; ,
\eeq
\beq
w_{D,E} \, = \, \int \, ds_1 \, ds_2 \, sign\left( s_1 - s_2 \right) \, W_{D,E} \; .  
\eeq
CLEO \cite{exp24} displays the forecast given by the GKS model and notice a slight 
discrepancy that shows up mainly in $w_A$. Then in order to have a better description
they modify the model by supplying 
some quantum-mechanical structure (a heritage of nuclear physics that accounts for the
finite size of hadrons) \cite{modcleo} that yields a good fit to data.
\par
Following the results of the EFT-like Lagrangian approach explained above, and once the
parameters are determined, it is possible to predict
the integrated structure functions.  By 
assuming isospin symmetry one can use the information obtained from the 
charged pions case to provide a description for the $\pi^- \pi^0 \pi^0$ final hadronic
state. The result and its comparison with the data is shown in Figure~\ref{fig:5}. For
$w_C$, $w_D$ and $w_E$, it can be seen that there is a good agreement in the low 
$Q^2$ region, while for increasing energy the experimental errors become too large
to state any conclusion (moreover, there seems to be a slight disagreement between
both experiments at several points). On the other hand, in the case of $w_A$ the 
theoretical curve seems to lie somewhat below the data for $Q^2 \gsim 1.5 \gev^2$.
However the study carried out in Ref.~\cite{us3pion} seems to conclude that this is 
due to some inconsistency between the data by CLEO and OPAL, on one side, and ALEPH
on the other.

\subsection{$\tau \rightarrow (K K \pi)^- \nu_{\tau}$}
These channels are more involved as both vector and axial-vector currents participate. 
Recently the CLEO Collaboration has published an analysis of the data collected on the 
$\tau^- \rightarrow K^+ K^- \pi^- \nu_{\tau}$ decay \cite{KLEO}. It was 
known that this process is not well described by the GKS model \cite{cleoprob} and therefore
in the new analysis they have reshaped the model with two new arbitrary parameters that
modulate both 
one of the axial-vector and the vector form factors (\ref{eq:fff}).
Afterward all the parameters are obtained through a reasonable fitting procedure.
Along these pages we have emphasized the fact that arbitrary parameterisations are of
little help in the procedure of obtaining information about non-perturbative QCD. In the CLEO example
just pointed out the new parameter in the vector form factor spoils the Wess-Zumino
anomaly normalisation, that appears at ${\cal O}(p^4)$ in $\chi PT$. It is true that there
are non-anomalous contributions proportional to the pseudoscalar masses at the next perturbative
order that could
account for a deviation but it would be surprising that the correction is around $80 \%$
as the fit points out. The real issue is that, as we have indicated, the GKS model is not
consistent with QCD and the CLEO reshaping is of not much use.
\par
Previous theoretical studies of these channels have only been done within parameterisations
in the line of the GKS model \cite{KS5,Gomez-Cadenas:1990uj}.
Recently we have employed the same technique than in the three pion case to study these
channels \cite{Pablo,usagain}. The main novelty has been to handle the vector form factor
through the construction of the relevant Lagrangian for the odd-intrinsic-parity sector, 
obtaining constraints on the new coupling constants through the requirement of the 
appropriate asymptotic behaviour of the vector form factor and also with some request to
the phenomenology \footnote{Unfortunately a comparison of our results with the experimental ones
\cite{KLEO} has not been possible because the experiment has not been able to provide us with
the data.}.
\par
Our analysis emphasizes the discrepancy between different approaches
in the weight of vector and axial-vector contribution to the decay width as can be seen in Table 1.

\begin{table}
\caption{Percentage of axial-vector and vector contribution to the total decay width
in the $\tau \rightarrow K K \pi \nu_{\tau}$ channel.}
\begin{center}
\vspace*{0.3cm}
\begin{tabular}{|ccc|}
\hline
\multicolumn{1}{|c}{Reference} &
\multicolumn{1}{c}{Axial-vector} &
\multicolumn{1}{c|}{Vector} \\
\hline
Our result \protect{\cite{Pablo}} &  $\sim 20 \% $  &  $\sim 80 \%$ \\
CLEO-III \protect{\cite{KLEO}} & $\sim 50 \%$ & $\sim 50 \%$ \\
\protect{\cite{Gomez-Cadenas:1990uj}} & $\sim 10 \%$ & $\sim 90 \%$ \\
\protect{\cite{KS5}} & $ \sim 60 \% $ & $ \sim 40 \% $ \\
\hline
\end{tabular} 
\end{center}
\vspace*{-0.7cm}
\end{table}

\section{Outlook}
Hadronic decays of the tau lepton provide all-important information on the hadronization
of currents in order to yield relevant knowledge on non-perturbative features of 
low-energy Quantum Chromodynamics. In order to achieve this goal we need to input more 
controlled QCD-based
modelizations. Our target is not only to fit the data at whatever cost, 
but do it with a reasonable parameterisation that allows us to understand more about the
theoretical description of Nature.
\par
The effective theory based phenomenological Lagrangian approach seems, along this line, more promising than the 
Breit-Wigner parameterisations. The procedure
relies in a field theory construction that embodies, up to a supposedly minor modelization of the 
large-$N_C$ behaviour, the relevant features of QCD in the resonance energy region, giving 
an appropriate account of the main traits of the experimental data and showing
that it is a compelling framework to work with. 

\vspace*{-0.1cm}
\section*{Acknowledgements}
I wish to thank Alberto Lusiani for his leading role in the excellent organization of the TAU06
meeting in Pisa (Italy) and for his patience with my delays. I also thank P.~Roig for a helpful reading
of the manuscript and W.~M.~Morse for pointing out an error in the first version of this preprint. 
This work has been supported in part
by the EU MRTN-CT-2006-035482 (FLAVIAnet), by MEC (Spain) under grant FPA2004-00996 and by 
Generalitat Valenciana under grants ACOMP06/098 and GV05/015.

\end{document}